# Application of Kalker's Theory of Rolling Contact for Dynamic Simulation of Mobile Robots

Jaswin[#1], Indrawanto[*2]

[#]Mechanical Engineering Department, Institut Teknologi Bandung
Jl. Ganesha No.10, Bandung, Indonesia
[1] jaswin_1998@yahoo.com
[2] indrawanto@ftmd.itb.ac.id

*Abstract*—**This paper presents a derivation of a dynamic simulation of a steerable-wheel mobile robot with wheel slip. The robot is controlled using pure pursuit algorithm. Kalker's simplified and linear theory of rolling contact are utilised to describe friction-creepage relationship between the robot wheel and the ground. Two simulations were created based on the two friction models. Simulation results of the two models are presented and compared.**

## I. INTRODUCTION

For mobile robots that experience high accelerations during their movement, dynamic simulations must include appropriate friction models as a function of wheel slippage. Existing robot simulations often use regularized Coulomb friction model. For example, [1] and [2] use a simplified function of wheel slip to describe the traction of robots with omnidirectional wheels. The numerical constants of these functions are determined experimentally on the physical mobile robots. Other works use empirical traction formula for pneumatic rubber tyres. For example, the work presented in [3] and [4] uses the Kiencke tyre model. The work in [5] uses the Magic formula. These empirical models for pneumatic tyres are not applicable for solid caster wheels used in many mobile robots, such as the robot modelled in this study.

The examples above indicate that there is a need for a more detailed analytical friction model for robot wheels. Moreover, an analytical model circumvents the need to perform experiments to determine numerical constants, which is an advantage during the design phase. Therefore, in this study, the theory of elastic rolling contact derived by J. J. Kalker was used to describe the dynamic equations of a steerable-wheel mobile robot. Kalker's theory was chosen because the assumptions used for its derivation, quasi-identity and half-space assumption, complies with the condition of rubber-solid ground contact [6].

The contents of this paper are: In section II, traction forces as a function of wheel creepages was derived using Kalker's simplified and linear theory of rolling contact. In section III, the dynamic equations of motion was determined. The final section contains the results of a dynamic simulation of the robot using pure pursuit algorithm.

It is hoped that this paper can offer an insight about the application of the theory of elastic bodies in rolling contact for dynamic simulation of mobile robots or other mechanical systems involving rolling motion.

## II. FRICTION MODEL

### A. Kinematics of Rolling Contact

When a wheel made of an elastic material rolls on another material, the velocity difference on the contact point of the two materials is called the slip velocity. There are two contributors to the slip velocity, that is, the rigid-body kinematics and the deformation of the material on the contact area.

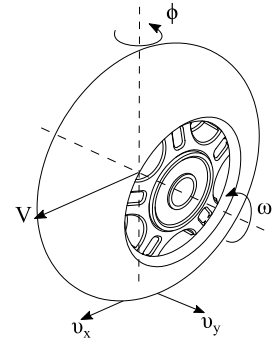

Fig. 1 Kinematics of a rolling wheel on flat ground

The rigid-body slip of the two contacting bodies is characterized by the creepage $\upsilon = (\upsilon_x, \upsilon_y)$ and spin $\phi$. The creepage and spin of a rolling body is defined by:

$$\upsilon_x = \frac{\Delta v_x}{V}, \upsilon_x = \frac{\Delta v_y}{V}, \phi = \frac{\Delta \omega_z}{V} \qquad (1)$$

Where $\Delta v_x$ and $\Delta v_y$ are the rigid body difference of velocity between the two contacting bodies, and V is the translational speed of the wheel. The variable $\upsilon_x$ is called longitudinal creepage, while $\upsilon_y$ is called the lateral creepage.

On the other hand, the elastic contribution to slip is caused by the compression or tension in the material on the contact area. The velocity associated with this deformation is given by:

$$\begin{aligned}\dot{u}_x &= -V\frac{\partial u_x}{\partial x} + \frac{\partial u_x}{\partial t} \\ \dot{u}_y &= -V\frac{\partial u_y}{\partial x} + \frac{\partial u_y}{\partial t}\end{aligned} \qquad (2)$$

Where u denotes the elastic deformation of the material in contact. If the rolling is assumed to have reached steady state, equation 2 gives:

$$\dot{u}_x = -V\frac{\partial u_x}{\partial x}$$
$$\dot{u}_y = -V\frac{\partial u_y}{\partial x} \quad (3)$$

For steady traction, the equation for total slip velocity $w_{x,y}$ in a point (x, y) on the contact area is a summation of equation 1 and 3. That is:

$$\frac{w_x}{V} = \upsilon_x - \phi y - \frac{\partial u_x}{\partial x}$$
$$\frac{w_y}{V} = \upsilon_y - \phi x - \frac{\partial u_y}{\partial x} \quad (4)$$

*B. Hertzian Contact*

Heinrich Hertz's contact theory describes the contact stresses that arise when two elastic bodies are pressed against each other. His theory was created to analyse the contact between glass lenses that were stacked together. In this work, this theory is used to determine the dimensions of the contact area. Hertz's theory follows several assumptions, that is:

- Homogenous and linear stress characteristic of the materials.
- Half-space assumption, that is, the contact area is much smaller than the bulk of the bodies.
- Constant curvature (radius) of the surfaces of the bodies near the contact area.
- Quasi-identity between the materials. Quasi-identity allows separate analysis of tangential and normal stresses.

The first three assumptions applies for the case of rubber caster wheels contacting solid ground. If the ground is also assumed to be perfectly rigid, and rubber is assumed to be incompressible, then the quasi-identity assumption is also valid [6].

According to Hertz's theory, when two elastic half-space bodies are pressed together with normal force N, their contact area is an ellipse with semi-axes a and b. The dimensions of the elliptical contact area are determined by a process shown in [7]. First, the curvatures of the bodies in contact, A and B, need to be determined. A and B depends on the principal radii of the bodies along the x and y axes. That is:

$$A = \frac{1}{2}(\frac{1}{R_{1y}} + \frac{1}{R_{2y}})$$
$$B = \frac{1}{2}(\frac{1}{R_{1x}} + \frac{1}{R_{2x}}) \quad (5)$$

In accordance to A and B, an intermediate variable θ is then calculated based on:

$$\cos\theta = \frac{|A-B|}{A+B} \quad (6)$$

From equation 5 and 6, the next step is to determine Hertzian constants λ, ν, and μ. These constants depend on θ, and can be found in tabulated form in [7]. They can then be used to calculate the length of the semi-axes of the contact area a and b, according to:

$$a = \lambda\left(\frac{3}{4}\frac{N}{E^*}\frac{1}{A+B}\right)^{1/3}$$
$$b = \nu\left(\frac{3}{4}\frac{N}{E^*}\frac{1}{A+B}\right)^{1/3} \quad (7)$$
$$\frac{1}{E^*} = \frac{1-\nu_1^2}{E_1} + \frac{1-\nu_2^2}{E_2}$$

Where $E_1, E_2$ are the elastic moduli of the materials in contact, and $\nu_1, \nu_2$ are the Poisson ratios.

*C. Kalker's Theory of Rolling Contact*

The next step is using Kalker's theory of rolling contact to find the forces acting on each wheel based on its lateral and longitudinal creepages. Kalker's theory is a variant of local Coulomb friction model, in which the friction forces working on a given point on the contact area is bounded by the normal pressure $p_z$ on that point multiplied by the coefficient of friction μ. The contact area is thus separated into adhesion area and slip area, depending on whether its friction has surpassed $p_z\mu$. This separation is illustrated in figure 2.

Two rolling contact theories are used in this study. They are the linear theory of rolling contact and the simplified theory of rolling contact [6]. The linear theory concerns cases where wheel creepage is small, such that the entire contact area is assumed to be in adhesion. That is, no slip occurs throughout the contact patch. In this case, longitudinal and lateral forces are given by a linear function of creepages:

$$F_x = -c^2 G C_{11} \upsilon_x$$
$$F_y = -c^2 G C_{22} \upsilon_y - c^3 G C_{23} \phi$$
$$M_z = -c^3 G C_{32} \upsilon_y - c^4 G C_{33} \phi \quad (8)$$
$$c = \sqrt{ab}$$

$C_{ij}$ are constants that are dependant on the combined Poisson ratio ν and the ratio of the semi-axes of the contact area. These constants are available in tabulated form in [6]. The constant G is called the combined modulus of rigidity, a function of the modulus of rigidity of the two materials. The combined Poisson ratio is given by:

$$\frac{\nu}{G} = \frac{1}{2}\left(\frac{\nu_1}{G_1} + \frac{\nu_2}{G_2}\right) \quad (9)$$

The constants of the linear theory is derived using an exact theoretical model. However, it is only valid for small creepages. To analyse cases where large creepages are present, the simplified theory may be used. The simplified theory arises from the assumption that the tangential stress **τ** = ($\tau_x$, $\tau_y$) on a point located at (x, y) on the contact area is proportional to its tangential displacement **u** = ($u_x$, $u_y$). That is:

$$\mathbf{u} = L\boldsymbol{\tau} \quad (10)$$

Proportionality constant L is derived from the constants of the linear theory (equation 8). This derivation is justified by the fact that in conditions where there is no slip area, the simplified theory must produce the same result with the linear theory.

$$L_x = \frac{8a}{3C_{11}G}, L_y = \frac{8a}{3C_{22}G}, L_\phi = \frac{\pi a\sqrt{a/b}}{4C_{13}G} \quad (11)$$

From equation 10, it follows that the traction of the wheel can be derived using equation 4, by substituting the elastic displacement $u_{x,y}$ with the tangential stresses $\tau_{x,y}$ multiplied by

$L_{x,y,\phi}$. In the stick region, total slip $w_{x,y}$ is equal to zero. Plugging into equation 4, we obtain:

$$\frac{\partial \tau_x}{\partial x} = \frac{v_x}{L_x} - \phi \frac{y}{L_\phi}$$
$$\frac{\partial \tau_y}{\partial x} = \frac{v_y}{L_y} - \phi \frac{x}{L_\phi} \quad (12)$$

When the traction becomes greater than the normal pressure times the coefficient of friction, sliding will occur and the total slip is not equal to zero. In that case:

$$|\tau| = p_z \mu \quad (13)$$

The direction of tangential stresses still follow equation 12, but the value follows equation 13.

The maximum value of tangential stress is determined from the distribution of normal pressure $p_z$ multiplied with the coefficient of friction μ. According to Hertz's theory, the distribution of $p_z$ is elliptical. However, following [8], a quadratic distribution is assumed in this study.

Based on equation 10-13, the calculation of total friction forces from creepages is done using an algorithm called FASTSIM [8]. This algorithm discretises the contact area into numerous segments, then perform numerical integration of the tangential stresses throughout the contact area to obtain total friction forces.

### D. Friction Model of Robot Wheel

In this section, the friction characteristics of the wheels used on the mobile robot is analysed using the linear theory and FASTSIM algorithm. The parameters of the robot wheel and the ground are given in table I. The ground is assumed to be perfectly rigid in comparison to the rubber wheels, such that its modulus of elasticity is taken as infinite.

TABLE I
WHEEL-GROUND PARAMETERS

| Parameters | Bodies | |
|---|---|---|
| | Wheel | Ground |
| Rx | 35 mm | ∞ |
| Ry | 12.5 mm | ∞ |
| $E_{1,2}$ | 6 MPa | ∞ |
| $G_{1,2}$ | 2 MPa | ∞ |
| $v_{1,2}$ | 0.5 | - |
| E* | 8 Mpa | 8 Mpa |
| G | 4 MPa | 4 MPa |
| v | 0.5 | 0.5 |
| μ | 0.7 | 0.7 |

Using the equations presented in section I, the Hertzian coefficient λ=0.95 and ν=0.72. Thus, the semi axes of the contact area is obtained as:

$$a = 1.14 \times 10^{-3} N^{\frac{1}{3}}$$
$$b = 8.64 \times 10^{-4} N^{\frac{1}{3}} \quad (14)$$

For example, if the ground reaction force is 60 N (a quarter of the weight of the robot), we obtain a = 4.46 mm and b = 3.38 mm.

Accordingly, the constants in equation 8 are shown in table II.

TABLE II
COEFFICIENTS OF ROLLING CONTACT

| Coefficient | Value |
|---|---|
| $C_{11}$ | 5.50 |
| $C_{22}$ | 4.53 |
| $C_{23}$ | 2.06 |

To perform analysis using the simplified theory, FASTSIM algorithm was implemented in Matlab. First, the contact area is discretised into 11 longitudinal slices. Then, those slices are separated further into 11 points. The algorithm then calculates the friction forces of each point. Finally, the total force is obtained by summing the local forces of all points. An example of the result of the algorithm is shown in figure 2 (N = 60 N, $v_x$ = 0.15, $v_y$ = 0.15).

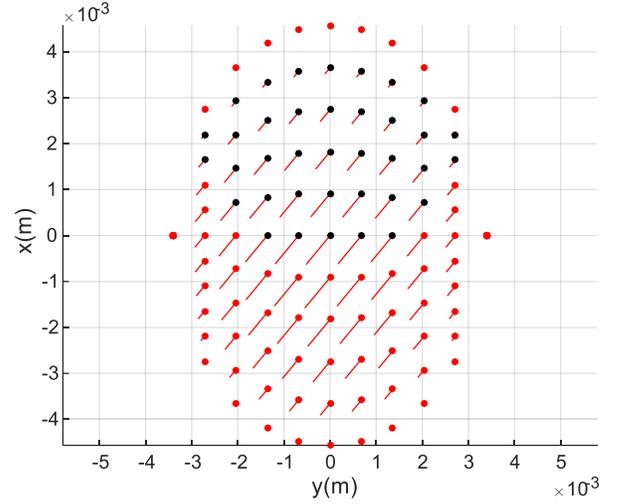

Fig. 2 Result of FASTSIM. Black points represent adhesion area, red points represent slip area. Red lines represent local friction forces.

The longitudinal friction coefficient as a function of longitudinal creepage is shown in figure 3. It is observed that for small creepages, the friction coefficient agrees with the linear theory. However, close to the full slip condition, the slope of friction-creepage curve becomes less steep. In other words, the linear theory will overestimate friction forces during the presence of large creepages.

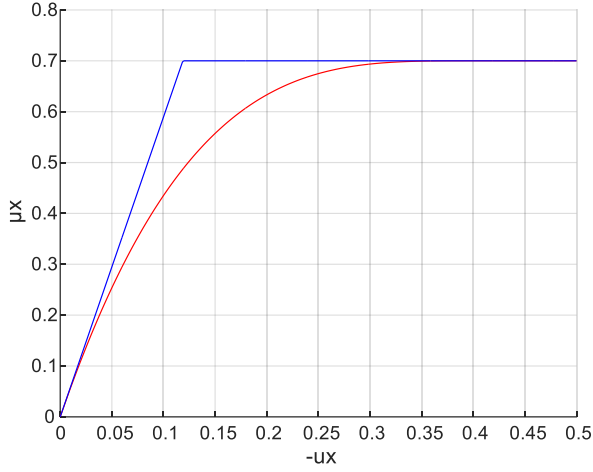

Fig. 3 Coefficient of friction as a function of longitudinal creepage (N = 60 N). Blue line calculated using the linear theory, red line calculated using the simplified theory.

Another characteristic of this friction model is that the coefficient of friction as a function of creepage is not constant for different normal forces. This is because the total friction force is proportional to the area of the contact patch, whereas the area is proportional to $N^{2/3}$. An illustration of this is shown in figure 4.

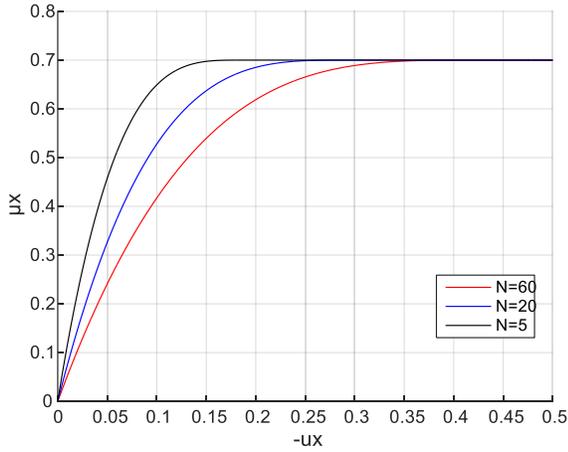

Fig. 4 Coefficient of friction as a function of longitudinal creepage for different normal forces.

## III. ROBOT MODEL

### A. Robot Kinematics

The kinematic equation of the robot follows figure 5. The robot has four wheels. Each of the wheels has a steering motor and can be independently steered to any direction.

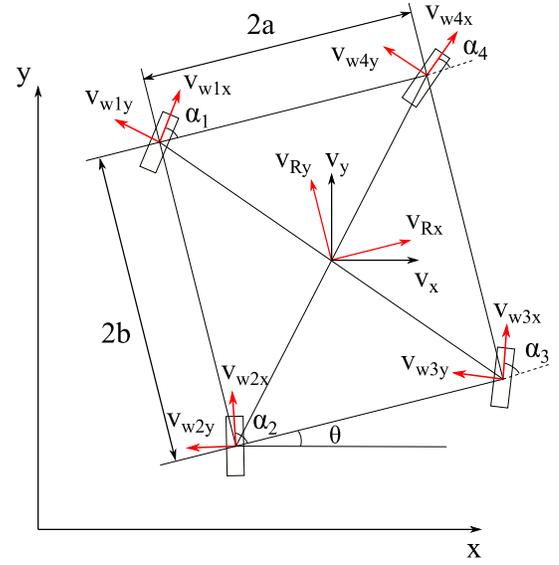

Fig. 5 Robot kinematics

The notation used for the kinematics of the robot is given by:
- $\mathbf{V}=(V_X\ V_Y\ 1)^T$: Robot linear velocity in global frame;
- $\mathbf{V_R}=(V_{Rx}\ V_{Ry}\ 1)^T$: Robot linear velocity in a frame attached to the robot;
- $\omega$: Angular velocity of the robot;
- $\alpha_i$: Steer angle of each robot wheel (i=1,2,3,4);
- $\mathbf{v_w} = (v_{wix}\ v_{wiy}\ 1)^T$: Velocity of each wheel along their longitudinal and lateral axes (i=1,2,3,4);
- $\omega_i$: Angular velocity of each wheel (i=1,2,3,4);
- r: Radius of the wheels;
- $\mathbf{r_i}$=position vector of wheel-i in the frame of the robot;

The velocity in the robot and global frame is related by:
$$\begin{bmatrix} V_{Rx} \\ V_{Ry} \\ 1 \end{bmatrix} = \begin{bmatrix} \cos\theta & \sin\theta & 0 \\ -\sin\theta & \cos\theta & 0 \\ 0 & 0 & 1 \end{bmatrix} \begin{bmatrix} V_x \\ V_y \\ 1 \end{bmatrix} \quad (15)$$

For each wheel-i, the velocity along its lateral and longitudinal direction follows:
$$\boldsymbol{v_{wi}} = \boldsymbol{R_{Wi}^R}[\boldsymbol{V_R} + \boldsymbol{\omega} \times \boldsymbol{r_i}]$$
$$\boldsymbol{R_{Wi}^R} = \begin{bmatrix} \cos\alpha_i & \sin\alpha_i & 0 \\ -\sin\alpha_i & \cos\alpha_i & 0 \\ 0 & 0 & 1 \end{bmatrix} \quad (16)$$

Using the notations above, the creepage of each wheel along their longitudinal and lateral axes are:
$$\begin{aligned} v_{ix} &= \frac{v_{wix} - \omega_i r}{v_{wix}} \\ v_{iy} &= \frac{v_{wiy}}{v_{wix}} \end{aligned} \quad (17)$$

## B. Robot Dynamics

The dynamic model of the robot is derived using figure 6. The centre of mass is assumed to be located on the robot geometric centre. The notations used are:
- $\mathbf{f_{wi}}=(f_{wix}\ f_{wiy})^T$: Lateral and longitudinal traction of wheel-i;
- m: Robot mass;
- I: moment of inertia of the robot around its centre of mass;

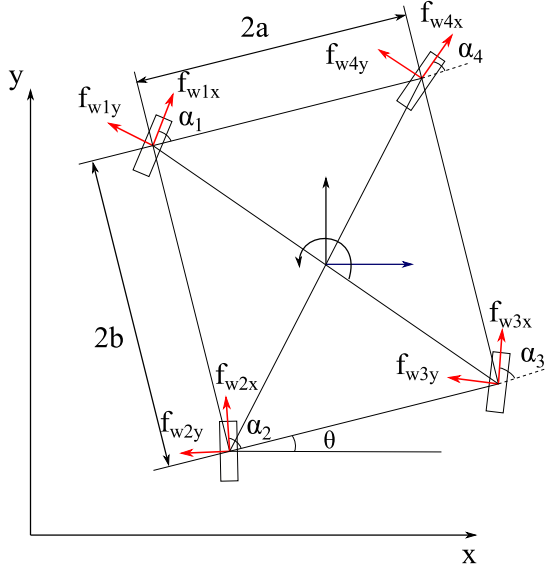

Fig. 6 Robot dynamics

From figure 7, the dynamic equations of the robot is derived using Newton-Euler equation of motion.

$$m\ddot{x} = \sum_{i=1}^{4} f_{wix} \cos \alpha_i + f_{wiy} \sin \alpha_i$$
$$m\ddot{y} = \sum_{i=1}^{4} f_{wix} \sin \alpha_i - f_{wiy} \cos \alpha_i \quad (18)$$
$$I\dot{\omega} = \sum_{i=1}^{4} \mathbf{r}_i \times \mathbf{f_{wi}}$$

## C. Calculation of Normal Forces

During its motion, the normal forces exerted by the ground on each of the robot wheels varies according to the acceleration of the robot. Using the static equilibrium equations of the robot, we obtain:

$$\begin{aligned} N_1 + N_2 + N_3 + N_4 &= mg \\ N_2 + N_3 - N_1 - N_4 &= ma_{YR}\frac{z}{b} \\ N_1 + N_2 - N_3 - N_4 &= ma_{xR}\frac{z}{a} \end{aligned} \quad (19)$$

Where $N_i$ denotes the normal force acting on wheel-i. Variable z denotes the height of the centre of mass. $a_{yR}$ and $a_{xR}$ are the inertial acceleration of the robot projected along the x and y axes of the robot coordinate frame. These accelerations are calculated from:

$$\begin{aligned} a_{xR} &= a_x \cos \theta + a_y \sin \theta \\ a_{yR} &= a_y \cos \theta - a_x \sin \theta \end{aligned} \quad (20)$$

Because the robot has four wheels, the normal forces are statically indeterminate. Therefore, it is additionally assumed that the normal force $N_i$ on each wheel depends linearly of its horizontal deflection $\delta_i$. This is similar to the method used in [3]. That is:

$$N_i = k\delta_i \quad (21)$$

Additionally, it is assumed that the robot base is perfectly rigid. Therefore:

$$\begin{aligned} \delta_1 + \delta_3 &= \delta_2 + \delta_4 \\ N_1 + N_3 &= N_2 + N_4 \end{aligned} \quad (22)$$

Using the equation 19 and 22, the normal force acting on each robot wheel can be obtained as:

$$\begin{aligned} N_1 &= \frac{mg}{4} + \frac{mza_{xR}}{4a} - \frac{mza_{yR}}{4b} \\ N_2 &= \frac{mg}{4} + \frac{mza_{xR}}{4a} + \frac{mza_{yR}}{4b} \\ N_3 &= \frac{mg}{4} - \frac{mza_{xR}}{4a} + \frac{mza_{yR}}{4b} \\ N_4 &= \frac{mg}{4} - \frac{mza_{xR}}{4a} - \frac{mza_{yR}}{4b} \end{aligned} \quad (23)$$

## D. Wheel Dynamics

The equation of motion for wheel–i of the robot is given by:
$$J\dot{\omega}_i = \tau_i - rf_{wix} \quad (24)$$
Where J is the moment of inertia of the wheel-motor assembly, r is the wheel radius, and τ is the torque of its driving motor. Only longitudinal traction affects the dynamics of the wheel.

The model of the DC motor used on the robot is:
$$V = K_M \omega + R\frac{\tau}{K_M} \quad (25)$$
Where V denotes the voltage applied to the motor, R is the resistance of the motor, and $K_M$ is the motor torque and back-emf constant.

The lateral and longitudinal traction depends on the creepage of the wheel at any given time. It is assumed that at all times, the traction distribution is steady. In order to determine traction forces, two friction models, model I and model II, was created. Model I uses the linear theory, and model II uses the simplified theory.

In order to find lateral and longitudinal forces, model I assumes the linear theory of rolling contact (equation 8) bounded by maximum coefficient of friction $\mu_{max}$. That is, defining:

$$\begin{aligned} F_{xh} &= -c^2 GC_{11}\upsilon_x \\ F_{yh} &= -c^2 GC_{22}\upsilon_y - c^3 GC_{23}\phi \\ F_h &= \sqrt{F_{xh}^2 + F_{yh}^2} \end{aligned} \quad (26)$$

- If $F_h < N\mu_{max} \rightarrow F_x = F_{xh}, F_y = F_{yh}$
- If $F_h \geq N\mu_{max} \rightarrow F_x = N\mu_{max}\frac{F_{xh}}{F_h}, F_y = N\mu_{max}\frac{F_{yh}}{F_h}$

Model II, on the other hand, uses FASTSIM to calculate total lateral and longitudinal friction forces. Model II is more accurate, especially at large creepage close to the full sliding condition. However, model II involves a longer calculation time, because of the numerical integration involved in FASTSIM.

The steering angle of each of the wheel is also controlled by a DC motor. The steering moment resulting from wheel spin is assumed to be small and is neglected. Ignoring steering moments caused by friction, the equation of motion for the steering system is:

$$V_S = K_{wS}\dot{\alpha}_i + \frac{R_S}{K_{TS}}J_S\ddot{\alpha}_i \qquad (27)$$

Where the subscript-s denotes the components of the steering system.

## IV. CONTROL SYSTEM

### A. Path Following using Pure Pursuit

The robot is controlled using pure pursuit method. Pure pursuit controls a robot so that it moves according to a given path. In pure pursuit, the robot directs its velocity vector towards the intersection of the path and a circle centred on the robot. This is illustrated in figure 7. The radius of the circle (called "pure pursuit radius") and the robot speed is manually tuned according to the given path.

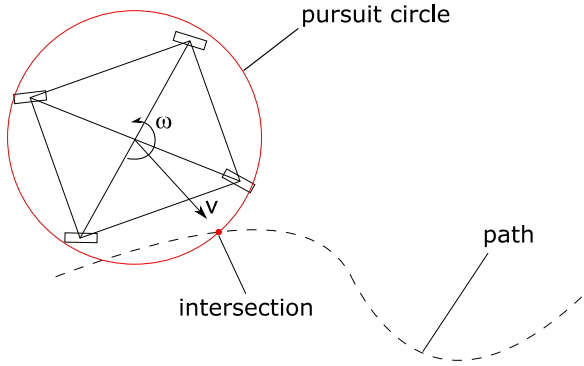

Fig. 7 Pure pursuit algorithm. The velocity is directed at the intersection of the pursuit circle and the desired path.

Using the pure pursuit method, it is possible to obtain magnitude and direction of the linear velocity of the robot. On the other hand, the heading of the robot is controlled using:

$$\omega = K_\omega(\theta_{ref} - \theta)$$

From v and ω, the angular velocity and steering angle of each wheel is obtain using the kinematic equations (equation 15, 16). The velocity of each wheel is then used as the reference of a low-level wheel velocity controller, using PID control. Similarly, the steering angle becomes the reference of the steering angle controller, using cascaded P-PI control. The overall control scheme of the robot is shown in figure 8.

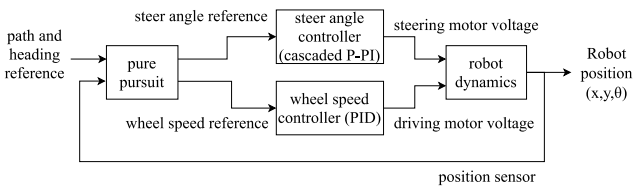

Fig. 8 Robot control system

## V. SIMULATION

In order to perform dynamic simulation, the parameters of the robot must first be determined. These parameters are obtained from the CAD model and the datasheet of the components used. The parameters of the robot are shown in table III.

TABLE IIIII
ROBOT PARAMETERS

| Parameter | Value |
|---|---|
| a | 0.4 m |
| b | 0.4 m |
| z | 0.2 m |
| m | 25 kg |
| I | 8 kgm^2 |
| J | 0.00032 kgm^2 |
| $J_S$ | 0.00375 kgm^2 |
| r | 35 mm |
| $K_M$ | 0.106 Nm/A |
| $K_{MS}$ | 0.833 Nm/A |
| R | 1.004 Ω |
| $R_S$ | 0.881 Ω |

Additionally, the battery used to power the robot has a maximum voltage of 24 V.

Based on table II and the equation of motion of the robot body and wheels, a dynamic simulation was developed. The algorithm of the dynamic simulation follows figure 9. The time-step of the simulation was set at 0.1 ms. The controller runs at a sampling time of 5 ms.

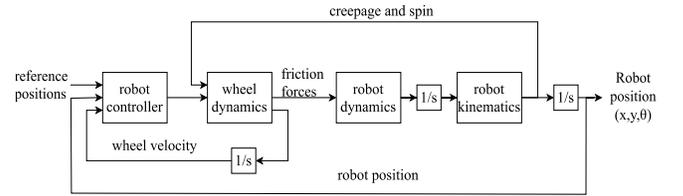

Fig. 9 Algorithm of dynamic simulation

For simulation, the robot is tasked with following a path shown in figure 10 using pure pursuit algorithm. The speed is set to be 2 m/s and the pursuit radius is set at 0.4 m. Resulting path of the robot is also plotted in figure 10. It can be seen that there is a noticeable difference between the paths found using the two friction models. The path calculated using FASTSIM (red) shows more deviation from the reference path. This is because as the total friction approaches the friction limit, the linear theory starts to overestimate the friction forces (see figure 4). Hence, it also underestimates the deviation of the robot from the reference path due to slip.

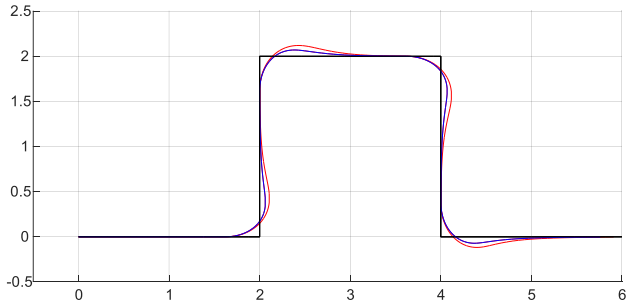

Fig. 10 Path of the robot during simulation. Blue path obtained from model I, red path from model II. Black line indicates reference path.

Figure 11 shows snapshots of the movement predicted using the simplified theory (model II).

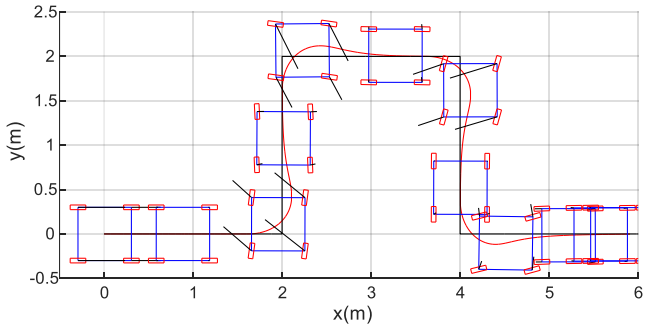

Fig. 11 Snapshots of robot during its motion (model II). Each snapshots are taken 0.5 s apart. The black lines indicates the total friction acting on each wheel.

Figure 11-16 shows the lateral, longitudinal, and normal friction forces acting on each wheel. Lateral friction spikes whenever the robot turns, whereas the longitudinal spike in friction occurs whenever the robot accelerates. The forces calculated using the linear theory is uniform for all wheels, whereas the simplified theory (model II) shows difference in lateral and longitudinal forces acting on each wheel at any given time.

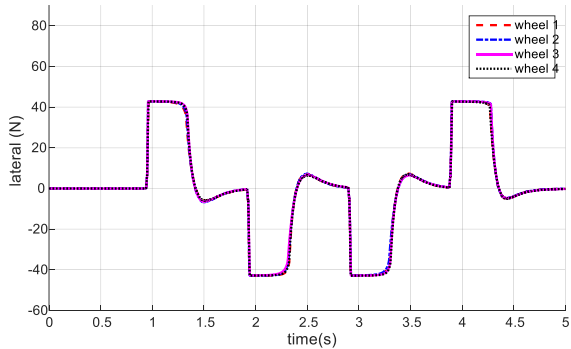

Fig. 11 Lateral forces on each wheel (model I)

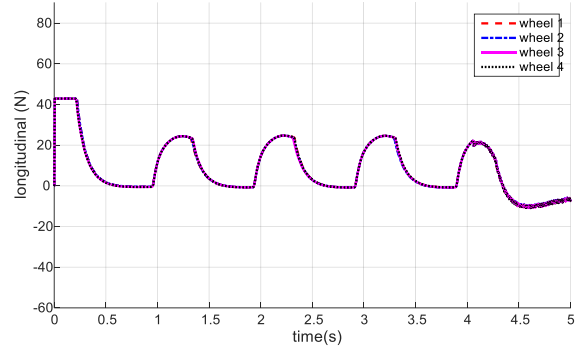

Fig. 12 Longitudinal forces on each wheel (model I)

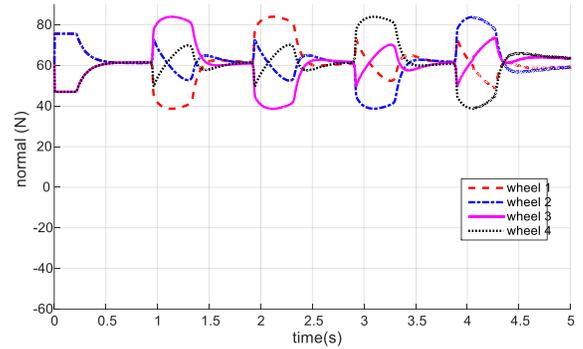

Fig. 13 Normal forces on each wheel (model I)

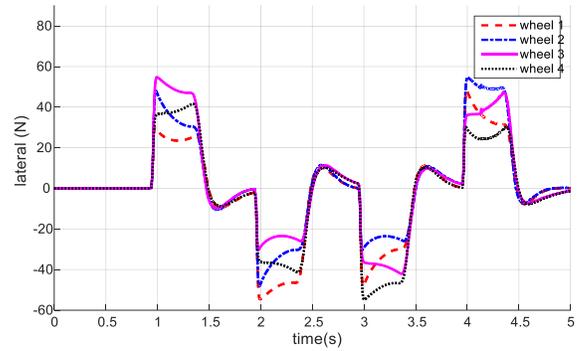

Fig. 14 Lateral forces on each wheel (model II)

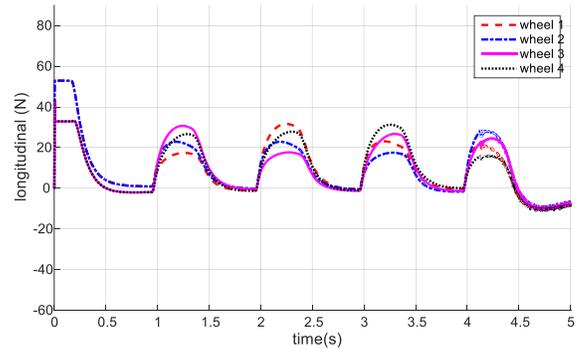

Fig. 15 Longitudinal forces on each wheel (model II)

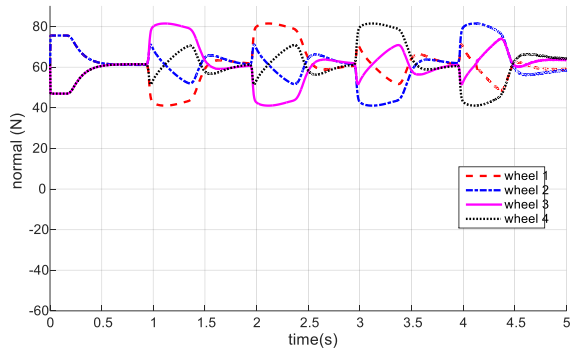

Fig. 16 Normal forces on each wheel (model II)

## VI. CONCLUSIONS

Kalker's theory of rolling contact has been used to derive the dynamic model of a steerable-wheel mobile robot. The robot was controlled using pure-pursuit algorithm. A dynamic simulation was then created based on the simplified and linear theory of rolling contact. It was observed that the two models give slightly different simulation results. The linear model allows faster computation compared to the simplified model. However, because the linear model overestimates traction for large creepages, the resulting path from the two friction models noticeably differ.